\documentclass[a4paper]{jpconf}

\usepackage{graphicx}
\usepackage{iopams}
\usepackage{bm}
\usepackage{subfigure}

\newcommand{\w}[1]{\bm{#1}}
\newcommand{\M}{{\mathcal M}}

\newcommand{\DSc}{\mathcal{D}}

\begin{document}

\title{Dynamical spacetimes and gravitational radiation in a Fully Constrained Formulation}

\author{Isabel Cordero-Carri\'on$^1$, Pablo Cerd\'a-Dur\'an$^2$ and Jos\'e Mar\'ia Ib\'a\~nez$^1$}

\address{$^1$ Departamento de Astronom\'ia y Astrof\'isica, Universidad de Valencia, C/ Dr. Moliner 50, E-46100 Burjassot, Valencia, Spain}
\address{$^2$ Max-Planck-Institut f\"ur Astrophysik, Karl-Schwarzschild-Strasse 1, D-85741 Garching, Germany}

\ead{isabel.cordero@uv.es, cerda@mpa-garching.mpg.de, jose.m.ibanez@uv.es}

\begin{abstract}
This contribution summarizes the recent work carried out to analyze the behavior of the hyperbolic sector of the Fully Constrained Formulation (FCF) derived in Bonazzola et al. 2004. The numerical experiments presented here allows one to be confident in the performances of the upgraded version of CoCoNuT's code by replacing the Conformally Flat Condition (CFC) approximation of the Einstein equations by the FCF.
\end{abstract}

\section{Introduction}
Most numerical codes used to solve the Einstein equations in order to obtain stationary or dynamical spacetimes generated by compact astrophysical objects are based on the 3+1 formalism (see, e.g., \cite{Alcubierre}). The Fully Constrained Formalism (FCF), recently proposed in \cite{Bonazzola04, CC08, CC09}, is a constrained evolution formulation of the full Einstein equations. It is a natural generalization of the Conformally Flat Condition (CFC) \cite{Isenberg, WilsoM89}, which is an approximation of the Einstein equations. FCF extends the CFC approximation and includes a hyperbolic system governing the gravitational radiation. Reader interested in details and motivation about FCF can address to the reference \cite{Bonazzola04}. We present numerical simulations of this system in several cases.

\section{Formalism}
Given an asymptotically flat spacetime $(\M, g_{\mu\nu})$ we consider a $ 3+1 $ splitting by spacelike hypersurfaces $\Sigma_t$, with $n^\mu$ being the timelike unit normal to $\Sigma_t$. Latin (greek) indices go from 1 (0) to 3.
$\gamma_{\mu\nu}= g_{\mu\nu}+n_\mu n_\nu$  denotes the 3-metric on $\Sigma_t$ and $K_{\mu\nu}=-\frac{1}{2}{\cal L}_{\w n}\gamma_{\mu\nu}$ the extrinsic curvature.
With the lapse function $N$ and the shift vector $\beta^i$, the metric $g_{\mu\nu}$ is expressed in coordinates $(x^\mu)$ as $g_{\mu\nu} \, dx^\mu \, dx^\nu =  - N^2 \, dt^2 + \gamma_{ij} (dx^i + \beta^i \, dt)(dx^j + \beta^j \, dt).$ As in~\cite{Bonazzola04}, we introduce a flat metric
$f_{ij}$, which satisfies $\partial_t f_{ij}=0$ and $f_{ij} \sim \gamma_{ij}$ at spatial infinity. We define $\gamma := \det \gamma_{ij}$ and $f := \det f_{ij}$. We introduce the following conformal decomposition: $\gamma_{ij} = \psi^4 \tilde{\gamma}_{ij}$. We define $h^{ij} := \tilde{\gamma}^{\;ij} - f^{ij}$. The gauge in~\cite{Bonazzola04} is \emph{maximal slicing}, $K=0$, and the so-called \emph{generalized Dirac gauge}, $\DSc_k \tilde{\gamma}^{\;ki} = 0$, where $\DSc_k$ is the Levi--Civita connection associated with $f_{ij}$. The Einstein equations become a coupled elliptic-hyperbolic system: the elliptic sector acts on the variables $\psi$, $N$, and $\beta^i$, and the hyperbolic sector acts on $h^{ij}$~\cite{Bonazzola04}. If $h^{ij}=0$ is imposed, CFC is recovered.

We introduce the conformal decomposition $\hat{A}^{ij} = \psi^{10} \left( K^{ij} - \frac{1}{3} K \gamma^{\;ij}\right)$, where $K=\gamma^{\;ij}K_{ij}$. This decomposition is different from the one introduced in~\cite{Bonazzola04}, but it is motivated by the local uniqueness properties of the elliptic equations shown in~\cite{CC09}. We rewrite the equation for $h^{ij}$ as a first order evolution system for the tensors $(h^{ij}, \hat{A}^{ij}, w^{ij}_k)$, where $w^{ij}_k := \DSc_k \tilde{\gamma}^{ij}$ \cite{CCphd}. 

\section{Numerical simulations}
We evolve numerically the previous evolution system for $(h^{ij}, \hat{A}^{ij}, w^{ij}_k)$. During the numerical simulations, we consider $N$, $\beta^i$, $\psi$ and the energy-momentum tensor as sources of the system. We perform the evolution of matter with the CoCoNuT code \cite{coconut}. Some basic elements of this code are: i) fourth order finite differences scheme for the spatial derivatives and fourth order Runge-Kutta methods for the time derivative; ii) axisymmetry and symmetry with respect to the equatorial plane; iii) spherical orthonormal coordinates, $(r,\theta, \varphi)$; iv) a Sommerfeld condition at the outer boundary; and v) Kreiss-Oliger dissipative term in order to avoid the numerical noise of high frequency which appears during long-term simulations.

\subsection{Teukolsky waves}
\begin{figure}
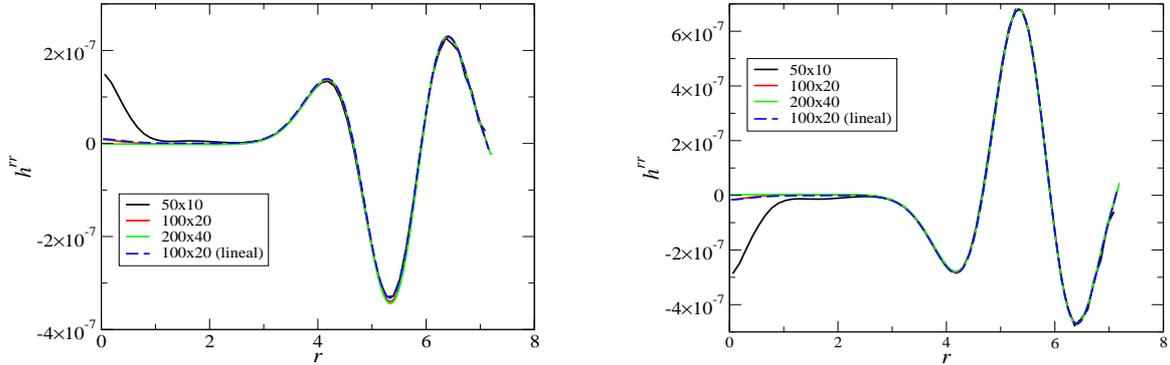

\begin{center}
\includegraphics[width=7cm,height=4.8cm]{hrr_e.eps} \hspace*{1cm}
\includegraphics[width=7cm,height=4.8cm]{hrr_p.eps}
\caption{Radial profile of $h^{rr}$ at $t=6$. Right figure corresponds to the equator and left figure corresponds to the pole. Continuous lines corresponds to $n_r=50$ and $n_{\theta}=100$ (radial and angular number of grid points), $n_r=100$ and $n_{\theta}=20$, and $n_r=200$ and $n_{\theta}=40$, in non-linear evolutions. Dashed line corresponds to $n_r=100$ and $n_{\theta}=20$, in a linear evolution.}
\label{fig:Teu_wave}
\end{center}
\end{figure}

The first test is the evolution of a combination of ingoing and outgoing even-parity axisymmetric Teukolsky waves~\cite{Teukolsky}, in order to construct regular initial data at the center $r=0$. The amplitude chosen is $10^{-5}$. These data are a solution of the linearized wave equation in a vacuum, they satisfy the Dirac gauge and are traceless (which is the linear approximation of unit determinant). The background is flat, i.e., $N = \psi = 1$, and $\beta^i = 0$. We display in Fig.~\ref{fig:Teu_wave} the radial profile of the component $h^{rr}$ at $t=6$, at the equator and at the pole, with different values of the number of radial and angular points respectively. The analytical solution is recovered in the linear regime, as well as the velocity and the amplitude of the wave, and its decay with the radius. The absolute errors (L2 norm) in the numerical simulations are smaller than $8\times10^{-8}$ for all the non zero components. The Sommerfeld condition at the outer boundary produces ingoing reflections that do not grow in time and have an amplitude much smaller than the amplitude of the initial one. We obtain second order of convergence for all the non zero components of the tensor $h^{ij}$, due to the influence of the inner boundary conditions imposed at $r = 0$, $\theta = 0$ and $\theta = \pi / 2$.

\subsection{Equilibrium configuration of rotating neutron stars}
We consider an axisymmetric and uniformly rotating neutron star in equilibrium. The initial data have been obtained from LORENE \cite{Lorene}, a library of spectral methods. These models, of different rotation parameters, have non-vanishing $\beta^i$, $N$, $\psi$ and matter fields. All these variables are kept fixed during the evolution of the tensor $h^{ij}$. The initial model is a neutron star with a 550 Hz rotation frequency, a  $1.6\;\mathrm{M}_{\odot}$ baryon mass and a 12.86 km coordinate equatorial radius.

\begin{figure}
\begin{center}
\includegraphics[width=13.5cm,height=5cm]{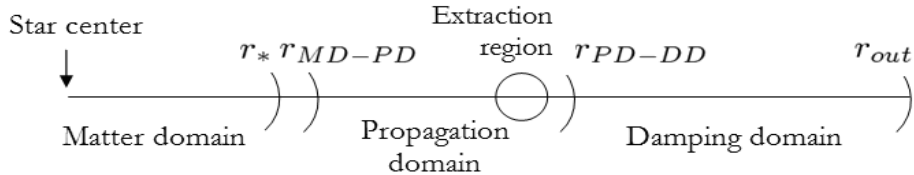}
\caption{Scheme of the radial grid used in the code for rotating neutron star simulations.}
\label{fig:grid}
\end{center}
\end{figure}

Figure~\ref{fig:grid} shows a sketch of the grid, where the following domains have been defined: The matter domain (MD) contains the star. The propagation domain (PD) is the one where the waves are well-resolved. The region for extracting the gravitational radiation is placed at the outer boundary of the PD. The damping domain (DD) has a lower resolution in the radial direction. The wave travels far enough for imposing the Sommerfeld condition at the outer boundary before reaching it. Let $r_{\mathrm{MD-PD}}$ be the radius which separates the MD and the PD, and $r_{\mathrm{PD-DD}}$ be the radius which separates the PD and the DD. The star radius, $r_*$, is close to th radius $r_{\mathrm{MD-PD}}$ but still smaller in value. The logarithmic scale (in the radial direction) is a very useful tool for constructing this kind of grid with a reasonable number of points. We get the necessary accuracy in the MD and the PD, but the time stepping becomes more severe than the one required in an equally spaced radial grid.

\begin{figure}
\begin{center}
\includegraphics[width=8cm,height=5.5cm]{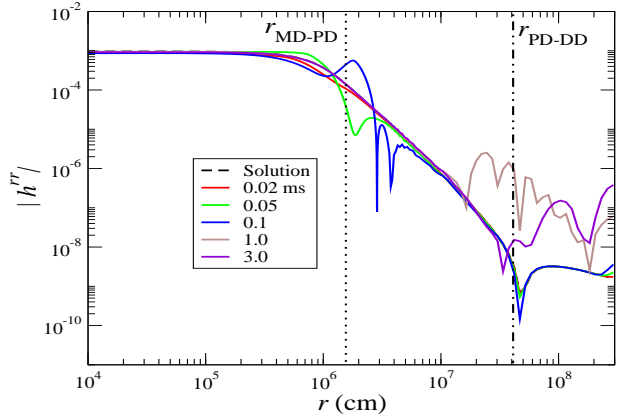}
\caption{Radial profile of $|h^{rr}|$. Dashed line corresponds to the stationary solution (as reference). Continuous lines correspond to the evolution of the initial data at 0.02, 0.05, 0.1, 1.0 and 3.0 ms respectively. Vertical lines denote radius between domains.}
\label{fig:stationarity}
\end{center}
\end{figure}

As initial data we have used stationary values for the tensor $h^{ij}$ and some differences from stationarity for $\hat{A}^{ij}$. The aim at introducing these initial differences from stationarity is two-fold: on the one hand, the recovering of stationarity from a perturbed initial data, and, on the other hand, testing the outer boundary Sommerfeld condition imposed by the generation of the artificial wave. In this general background, those differences from stationarity introduced in the initial data, generate a perturbation that propagates to the outer boundary, leaving behind the stationary solution. In Fig.~\ref{fig:stationarity}, we have plotted the absolute value of the component $h^{rr}$ in terms of the radius for different times. We can see the perturbation traveling towards the outer boundary, where the wave leaves the numerical domain. The differences from stationarity are interpreted as an initial perturbation. The reflections close to the outer boundary come from the numerical implementation of the Sommerfeld condition. Second order of convergence is obtained for all the non zero components of the tensor $h^{ij}$, as previously.

\begin{figure}
\begin{center}
\vspace{1cm}
\includegraphics[width=8cm,height=5.5cm]{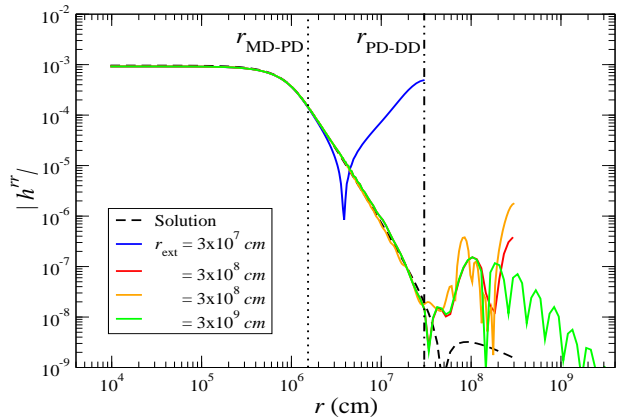}
\caption{Radial profile of $|h^{rr}|$ at time $t=3$ ms. The dashed line corresponds to the stationary solution (as reference). Continuous lines correspond to the numerical evolutions whose outer boundary are placed at $3 \times 10^7$, $3 \times 10^8$ and $3 \times 10^9$ cm. Vertical lines denote radii between domains.}
\label{fig:out_bound}
\end{center}
\end{figure}

This test helps to understand how the Sommerfeld condition at the outer boundary works. This condition is imposed to treat the outgoing waves which reach the outer boundary, as the one caused by the initial data. If the background is not flat enough, the Sommerfeld condition can interpret the background as an outgoing wave. This behavior can be seen in Fig.~\ref{fig:out_bound} for the simulation whose outer boundary is placed closer, at $3 \times 10^7$. The Sommerfeld condition works properly for $r_{out} \gtrsim 3 \times 10^8$cm (around 300 stellar radii). In our calculations we have used 80 radial grid points in the MD, 80 points in the PD and around 50 in the DD.

\begin{figure}
\begin{center}
\includegraphics[width=8cm,height=5.2cm]{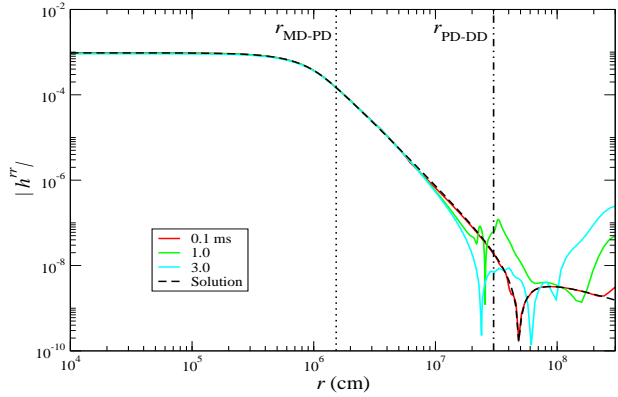}
\caption{Radial profile of $|h^{rr}|$. The dashed line corresponds to the stationary solution (as reference). Continuous lines correspond to the evolution of the initial data at 0.1, 1.0 and 3.0 ms. Vertical lines denote radii between domains.}
\label{fig:sta_ini_data}
\end{center}
\end{figure}

Once the outer boundary radius has been fixed, we can use the stationary initial data for both $h^{ij}$ and $\hat{A}^{ij}$ tensors. In Fig.~\ref{fig:sta_ini_data} we show the absolute value of the component $h^{rr}$, in terms of the radius, for different times. No initial artificial wave is introduced. The noise comes from the outer boundary condition. Again, second order of convergence is obtained.

\subsection{Perturbed equilibrium configuration of a rotating neutron star}
We consider a perturbed neutron star. The perturbation is proportional to $\sin(\pi\;r/r_*)$. Since the star is rotating, $r_*$ depends on $\theta$, and the perturbation is not spherically symmetric; hence, gravitational radiation is expected to be generated. The hydrodynamic equations governing the matter evolution are solved. We keep fixed $N$, $\beta^i$ and $\psi$, assuming that they do not change very much during the evolution. This test will provide where to place $r_{\mathrm{PD-DD}}$, and, consequently, where to extract the gravitational radiation.

\begin{figure}
\begin{center}
\vspace{0.2cm}
\includegraphics[width=8cm,height=4.8cm]{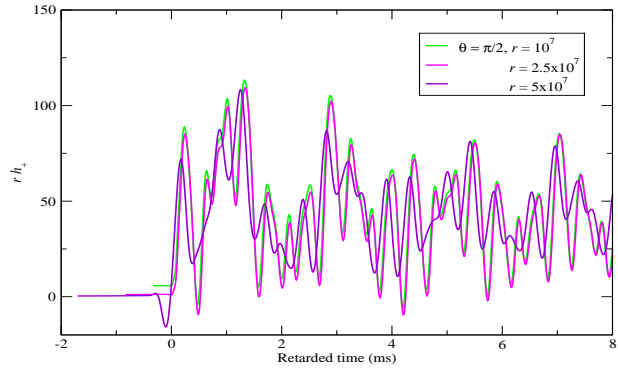}
\caption{$rh_+$ versus the retarded time. The corresponding wave is extracted at $r=10^7$, $r=2.5 \times 10^7$ and $r=5 \times 10^7$cm, at the equator. $r_{\mathrm{PD-DD}}$ is placed at $3 \times 10^7$cm.}
\label{fig:wave_e}
\end{center}
\end{figure}

In Fig.~\ref{fig:wave_e}, an approximation of the real part of the Weyl scalar $\Psi_4$, $h_+$, (scaled with $r$) is plotted in terms of the retarded time, for different values of the extraction radius, at the equator. This approximation is accurate enough for the objectives of this test. $r_{\mathrm{PD-DD}}$ is placed at $3 \times 10^7$cm (around 30 stellar radii). The waves correspond to the physical gravitational waves coming from the  evolution of the perturbed star. This evolution is governed by the coupled hydrodynamic and the Einstein equations (but keeping fixed $N$, $\beta^i$ and $\psi$). The different curves in Fig.~\ref{fig:wave_e} refer to different radii where the gravitational radiation has been extracted. From this figure, we can deduce that the speed of the waves is the light velocity and they decay as $1/r$. We also see that the wave must be extracted far from the source but inside the PD. Regarding the resolution of the logarithmic grid, we have noticed that it is enough to cover a wavelength with five points at the end of the PD in order to obtain the wave accurately.
 
\begin{figure}
\begin{center}
\includegraphics[width=8cm,height=4.8cm]{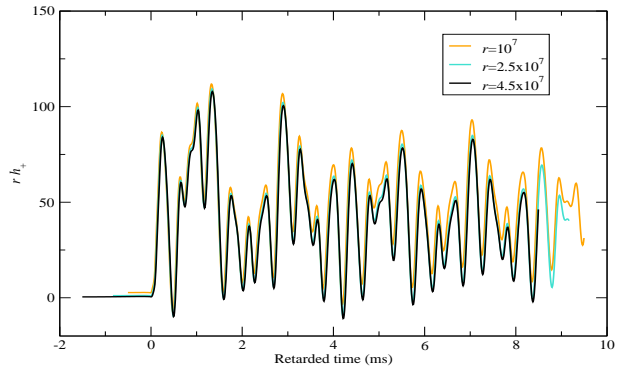}
\caption{$rh_+$ versus the retarded time, at the equator. Radii are expressed in cm. The corresponding wave is extracted at $r=10^7$ ($r_{\mathrm{PD-DD}}=1.5 \times 10^7$), $r=2.5 \times 10^7$ ($r_{\mathrm{PD-DD}}=3 \times 10^7$) and $r=4.5 \times 10^7$ ($r_{\mathrm{PD-DD}}=5 \times 10^7$).}
\label{fig:extrac_r}
\end{center}
\end{figure}

Furthermore, it will be important to know where to place $r_{\mathrm{PD-DD}}$. In Fig.~\ref{fig:extrac_r}, $rh_+$ is plotted in terms of the retarded time in simulations with different $r_{\mathrm{PD-DD}}$. The wave is extracted close to $r_{\mathrm{PD-DD}}$ in all the cases. From the figure, it is reasonable to place $r_{\mathrm{PD-DD}}$ at $3 \times 10^7$cm (around 30 stellar radii) or further.

\section{Conclusions}
The first version of CoCoNuT's code \cite{Dimmelmeier02} was designed to evolve matter fields of a perfect fluid in the dynamical spacetime of the CFC approximation. The current version of CoCoNuT incorporates magnetic fields, i.e., it is a general-relativistic magneto-hydrodynamic code which evolves matter in the dynamical spacetime of CFC. The present contribution to this conference, together with previous works in \cite{CC08, CC09}, is a step towards the upgrading of the metric evolution of CoCoNuT's code by substituting the CFC approximation of the Einstein equations by the FCF of \cite{Bonazzola04}. The hyperbolic sector to evolve $h^{ij}$ has already been included into CoCoNuT's code. In this contribution we have shown that this sector works properly. We have presented the numerical evolution of Teukolsky waves in a flat background. We have analyzed the behavior of the 
code under the Sommerfeld boundary condition, and detected the outer radius, in the case of spacetimes generated by equilibrium configurations of rotating neutron stars. Finally, we have extracted the gravitational waveforms in the case of dynamical spacetimes generated by perturbed rotating neutron stars.

\ack
I. C.-C. acknowledges support from the Spanish Ministerio de Educaci\'on y Ciencia (AP2005-2857). This work has been supported by the Ministerio de Educaci\'on y Ciencia through Grant No. AYA2007-67626-C03-01 and by the Collaborative Research Center on Gravitational Wave Astronomy of the Deutsche Forschungsgemeinschaft (DFG SFB/Transregio 7).

\end{document}